\documentstyle[aps,epsfig,axodraw,preprint]{revtex}
\tighten
\begin{document}
\title{A Numerical Experiment in DLCQ: Microcausality, Continuum Limit and
all that}
\author{{\bf Dipankar Chakrabarti, 
 Asmita Mukherjee{\thanks{email: asmita@tnp.saha.ernet.in}}, Rajen Kundu, A. Harindranath$^{}$} \\
	$^{}$Saha Institute of Nuclear Physics\\ 1/AF Bidhan Nagar, 
		Calcutta, 700064 India} 
\date{February 2, 2000}
\maketitle
\begin{abstract}
Issues related with microcausality
violation and continuum limit in the context of (1+1) dimensional scalar
field theory in discretized
light-cone quantization (DLCQ) are addressed in parallel with 
discretized equal time quantization (DETQ) and the fact that Lorentz
invariance and microcausality are restored if one can take the continuum
limit properly is emphasized. 
In the free case, it is shown with numerical evidence that 
the continuum results can be reproduced from DLCQ results for 
the Pauli-Jordan function and the real part of Feynman propagator.  
The contributions coming from $k^+$ near zero  region 
in these cases are found to be very small.
In the interacting case, aspects related to the continuum
limit of DLCQ results in perturbation theory in momentum space are discussed. 
\end{abstract}
\vskip .4in
%\centerline{PACS: 11.10.Ef, 11.10.Gh, 11.55.Hx, 13.40.-f}
%\vskip .2in
\newpage
\noindent {\bf 1. Introduction}
\vskip .2in
%%%%%%%%%%%%%%%%%%%%%%%%%%%%%%%%%%%%%%%%%%%%%%%%%%%%%%%%%%%%%%%%%%%%%%%%%%%
The discretized light-cone quantization (DLCQ) \cite{mas,cas,tho,pau} 
was proposed to study
nonperturbative aspects of field theories and is extensively used in practical
calculations following Ref. \cite{pau}.  The vacuum is simpler 
 and the treatment of the infrared degrees of freedom is less complicated in
this formalism. DLCQ has also been applied in different contexts such as
transverse lattice formalism \cite{sande} and M-theory \cite{sus}. However,
at the same time it has some inherent difficulties \cite{yam}.        

In a recent work\cite{sch} it is shown that microcausality is violated in
the framework of DLCQ.
 The results of \cite{sch} can be summarized as follows. Consider the Pauli-Jordan function 
in (1+1) dimensional free scalar theory
\begin{eqnarray}
\Delta(x)~=~{1\over i}~[\phi(x),\phi(0)]~=~
{1\over i}~\int~ {d^2k\over 2\pi} \delta(k^2-m^2) \epsilon(x^0)
e^{~-ik\cdot x}\, .
\end{eqnarray}
An explicit evaluation leads to,
\begin{eqnarray}
\Delta(x)~=~-{1\over 2}\epsilon(x^0)\theta(x^2)J_0(m\sqrt{x^2})~
=~-{1\over 4}[\epsilon(x^+)~+~\epsilon(x^-)]~J_0(m\sqrt{x^2})\, ,
\label{pjd}
\end{eqnarray}
where $J_0$ is the Bessel function and the light-front variables are defined as 
$x^{\pm}=x^0\pm x^1$.
As is evident from Eq. (\ref{pjd}), $\Delta(x)$ is zero in the spacelike
region due to microcausality. Numerical results show that 
 $\Delta(x)$ is non-vanishing in the space-like region
within the box if the box is considered to be lying along light-front
longitudinal direction $x^-$ (i.e., in DLCQ) and  
hence violates microcausality. Further, the continuum result
for $\Delta(x)$ is not restored if one uses DLCQ and let $L \rightarrow
\infty$. Similar observation was
also made in \cite{sal}. This is in contrast to the
case when one uses the box to be lying along $x^1$ (i.e., in DETQ) where
 microcausality is not violated within the box as long as
$x^0<L$ and continuum limit is reproduced. 

In view of the fact that DLCQ is already
known to produce reasonable results\cite{brodsky}, it is necessary to
investigate further to clarify what actually is going on. In our opinion,
the observation made in Ref.\cite{sch} is, at best, superficial as we 
have clarified below providing the actual picture with numerical evidence. 
\vskip .2in
%%%%%%%%%%%%%%%%%%%%%%%%%%%%%%%%%%%%%%%%%%%%%%
\noindent{\bf 2. Comments on Microcausality}
%%%%%%%%%%%%%%%%%%%%%%%%%%%%%%%%%%%%%%%%%%%%%%
\vskip .2in
First of all, it should be noted that when one is using DLCQ or DETQ, one is
dealing with a theory where the raw output is not the ultimate concern. Only
the continuum limit of that output is what we are interested in and
presumably can be tested with some experiment. In this sense, it is not our
primary concern if some symmetry like boost invariance or even
microcausality is violated in the discretized version of the theory. The fact
that microcausality may be 
violated in DLCQ {\it with a fixed box length} $2L$, 
can be seen trivially once we assume the periodic boundary condition
$\phi(L)=\phi(-L)$, since
\begin{eqnarray}
~~~[\phi(L),\phi(0)]~=~[\phi(-L),\phi(0)]~=~[\phi(0),\phi(L)], ~~~~~
\Rightarrow~~~ [\phi(L),\phi(0)]=0\, ,
\end{eqnarray} 
for all time. If the continuum result is our guideline, we should have
obtained $-{i\over 4}\epsilon(L)$ for $x^+=0$ and hence the microcausality that
is always maintained in the continuum theory is not guaranteed to be
respected in DLCQ {\it with fixed $L$}, as also is the case with 
boost invariance. That precisely is the reason for
which one has to remove the $L$-dependence of the result obtained from 
discretized version of the theory (which, in general, may depend on $L$) by
taking continuum limit to ensure correct physics. Therefore, if the
continuum result for Pauli-Jordan function is not reproduced from DLCQ as is
claimed in Ref.\cite{sch}, we are facing a real disaster. In the rest of this
work, we show that the situation is not so alarming and try to clarify
the intricacy involved in taking the continuum limit, both in DLCQ and DETQ.
We also compare and contrast the situations in coordinate space and momentum
space.
\vskip .2in
%%%%%%%%%%%%%%%%%%%%%%%%%%%%%%%%
\noindent{\bf 3. Comments on $k^+=0$ modes}
%%%%%%%%%%%%%%%%%%%%%%%%%%%%%%%%
\vskip .2in
The Pauli-Jordan function can be given in the momentum representation by
integrating $k^0$ or $k^-$ as
\begin{eqnarray}
\Delta(x)&=& -~\int_{-\infty}^\infty {dk^1\over 2\pi \omega_k }~\sin(\omega_k x^0 ~-~k^1
x^1)\, ,\nonumber\\
&=&-~\int_0^\infty {dk^+\over 2\pi k^+ }~\sin(k^- x^+/2~+ ~k^+
x^-/2)\, ,\label{clc}
\end{eqnarray}
where $\omega_k=\sqrt{(k^1)^2+m^2}$ and $k^-=m^2/k^+$. 

Notice that the integrand in Eq. (\ref{clc}) is highly oscillatory for $k^+$
near $0$ in contrast
to the corresponding equal-time case when $k^1$ near $0$ as long as the mass is 
not
zero. In general, understanding the role of $k^+=0$ modes is very
crucial in light-front quantization\cite{yam}.   
The observed discrepancy between DLCQ result
and the continuum one for $\Delta(x)$ is generally believed to be due to  
the elimination of the zero mode ($k^+=0$, the accumulation point) 
in DLCQ , at least in the case of
free theory\cite{sch,sal}. To see the actual contribution to 
$\Delta(x)$,  
we have performed numerically the
integration of Eq. (\ref{clc}) separately for different $k^+$ regions and the
results are shown in Fig. 1. It is clear that the contribution coming from
the region where $k^+$ is very small is negligible compared to that coming 
from the other region. 
\vspace{.5cm}
\begin{center}
\epsfig{figure=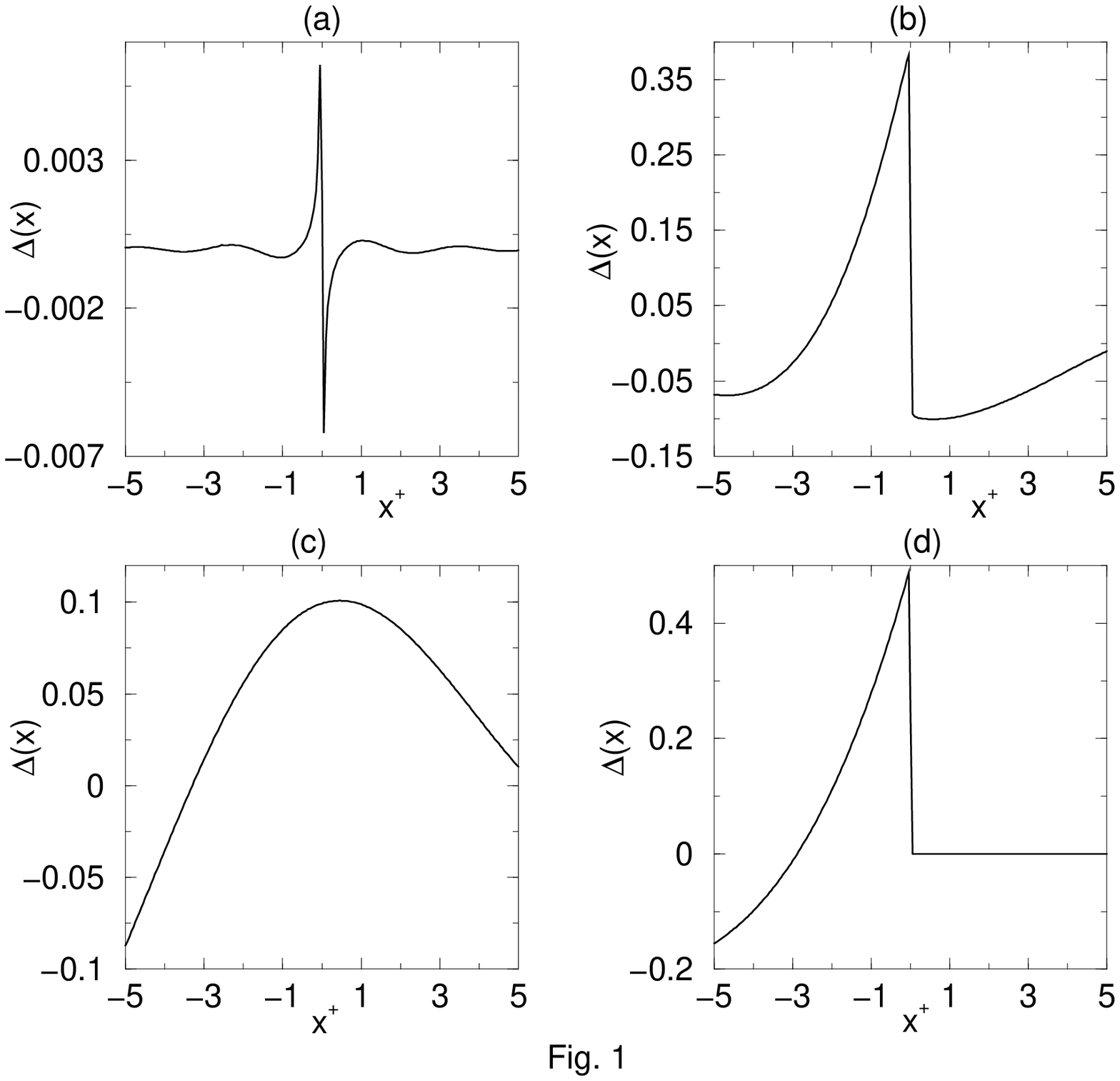,width=10.0cm,height=10.0cm}\\
\end{center}
\vspace{0.2cm}
\begin{center}
\parbox{14.0cm}
{{\footnotesize 
 Fig. 1:  Plot of Pauli-Jordan function showing contributions from
different regions of $k^+$ 
 for $x^-=-2.0$ and $m^2=1$.
 The ranges of $k^+$ integrations are (a) $10^{-5}$ 
to $10^{-3}$, (b) $ 10^{-3}$ to 1.0, (c) 1.0 to $10^3$ and (d) $10^{-5}$ to
 $10^3$, this curve is the same as that of Eq. (\ref{pjd}).}}
\end{center}
\vspace{0.3cm}

Though it is not possible to reach
$k^+=0$ numerically, it can be seen that the total result matches with the
exact result (Eq. (\ref{pjd})) suggesting that the contribution from 
the region
$k^+\le 10^{-5}$ is even smaller and insignificant. 
So  the
discrepancy does  not seem to be caused by the zero modes. Our numerical
result in the continuum agrees with the assertion in Ref.\cite{yam}, that
zero mode is not an accumulation point for the Pauli-Jordan function. For more
discussions on the role of zero modes see Ref.\cite{mas,yam}.
%%%%%%%%%%%%%%%%%%%%%%%%%%%%%%%%%%%%%%%%%%%%%%%%%%%%%%%%%%%%%%%%%%%%
\vskip .2in
\noindent{\bf 4. Continuum Limit}
\vskip .2in
%%%%%%%%%%%%%%%%%%%%%%%%%%%%%%%%%%%%%%%%%%%%%%%%%%%%%%%%%%%%%%%%%%%%      
The discretized
versions
of $\Delta(x)$ by restricting spatial coordinates, $-L\leq x^1,x^-\leq L$
and using periodic boundary condition for the field $\phi$ are given by,
\begin{eqnarray}
&&\Delta_{ET}(x)=-\sum_{n=-N}^{N}~{1\over 2\omega_n L}
\sin(\omega_n x^0 -n\pi x^1/L)\, ,\label{det}\\
&&\Delta_{LC}(x)=-\sum_{n=1}^{N}~{1\over 2\pi n }\sin(k^-_n x^+/2 +n\pi
x^-/L)\, ,\label{dlc}
\end{eqnarray}
where, in principle, $N\rightarrow\infty$ and for $n$th discrete
momentum mode energies are given by $\omega_n=\sqrt{n^2\pi^2/L^2+m^2}$ 
and $k^-_n=m^2L/2\pi n$ respectively. 

In general, at the end of 
any calculation in the discretized version of the theory one is
expected to get back the continuum result by taking $N\rightarrow\infty$
and $L\rightarrow\infty$ limit. For practical computations,
 it means that one has to take $N$ sufficiently large
 for any particular  $L$ value such that the summation saturates
(in the sense that it does not change appreciably if the value of $N$
is increased). It turns out that if $N/L$ ratio is large then the 
saturation occurs. In all the results presented $N$ has been taken
sufficiently large so that the results are independent of $N$. 

 Now, in DLCQ in coordinate space, taking the continuum limit is
a little tricky. To understand things properly, we perform the summation in
Eq. (\ref{det}) and Eq. (\ref{dlc}) numerically and study their behavior
by changing the  parameters involved, namely, $N$, $L$ and also the
space-time points at which it is calculated.

First, we present the observation made in DETQ. In Fig. 2(a), we have plotted
$\Delta_{ET}$ in the continuum theory. From Fig. 2(b),  
we can clearly see that for a
particular $L(>x^0)$ and for fixed $x^0$, $\Delta_{ET}$ actually oscillates about
the continuum value and {\it converges} 
to it (at least within the box) as we 
increase $N$ sufficiently. {\it Since continuum result is reproduced within
the box, we see that microcausality is not violated within the box}, i.e.,
yields zero for spacelike region.  
Outside the box it is a different story altogether, since we just get 
periodic copies which, of course, differs from the continuum result. Now, 
increasing
only $L$ with $N$ fixed to earlier value, the result again starts
oscillating around the continuum value and one could even get widely varying
result (non-zero for spacelike region) if $L$ is sufficiently 
increased as shown in Fig. 2(c) (periodic copies are not shown there). 

\vspace{0.5cm}
\begin{center}
\hspace{-0.0cm}
\parbox{5cm}{\epsfig{figure=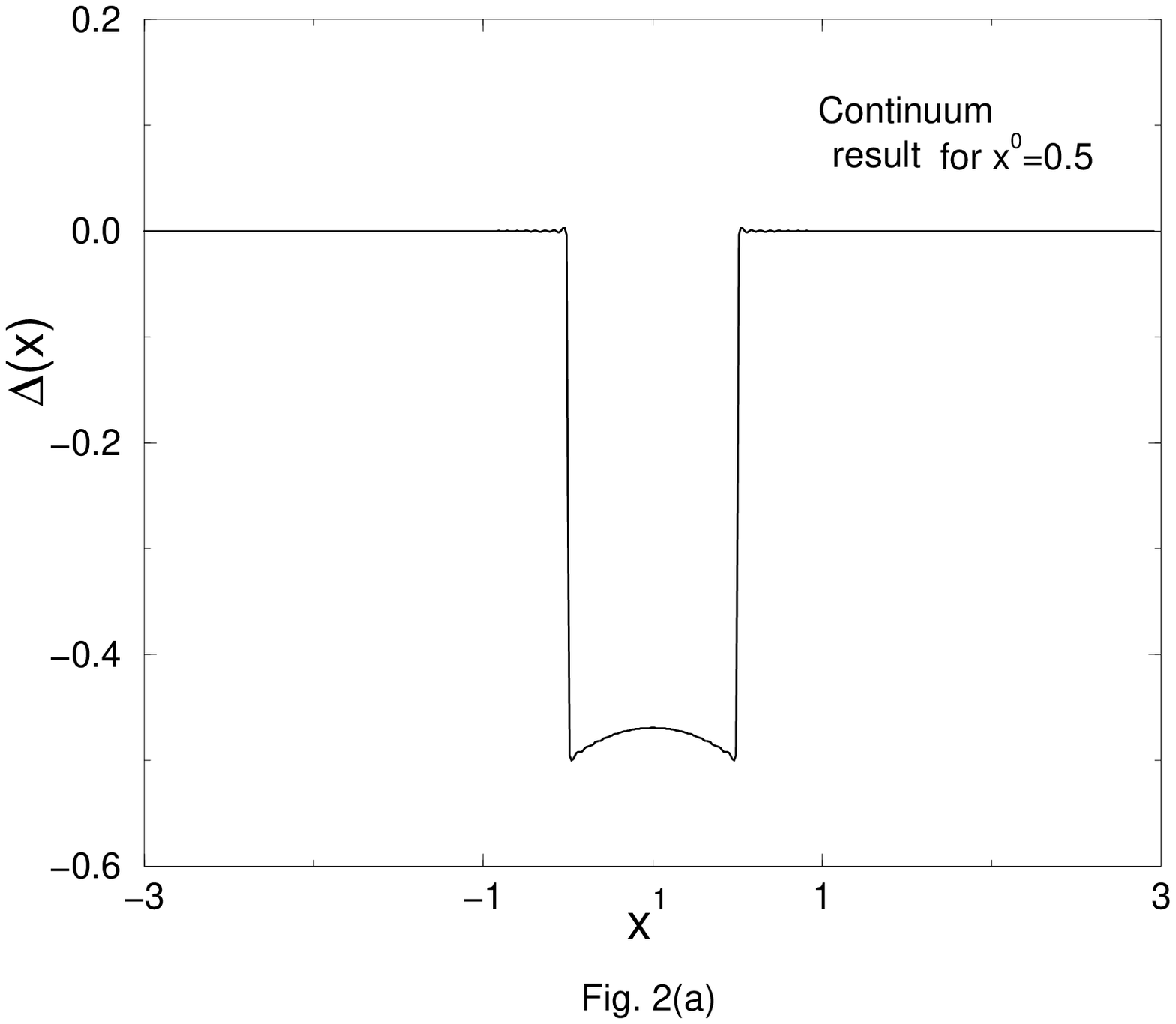,width=5cm,height=7cm}}\ \
\parbox{5cm}{\epsfig{figure=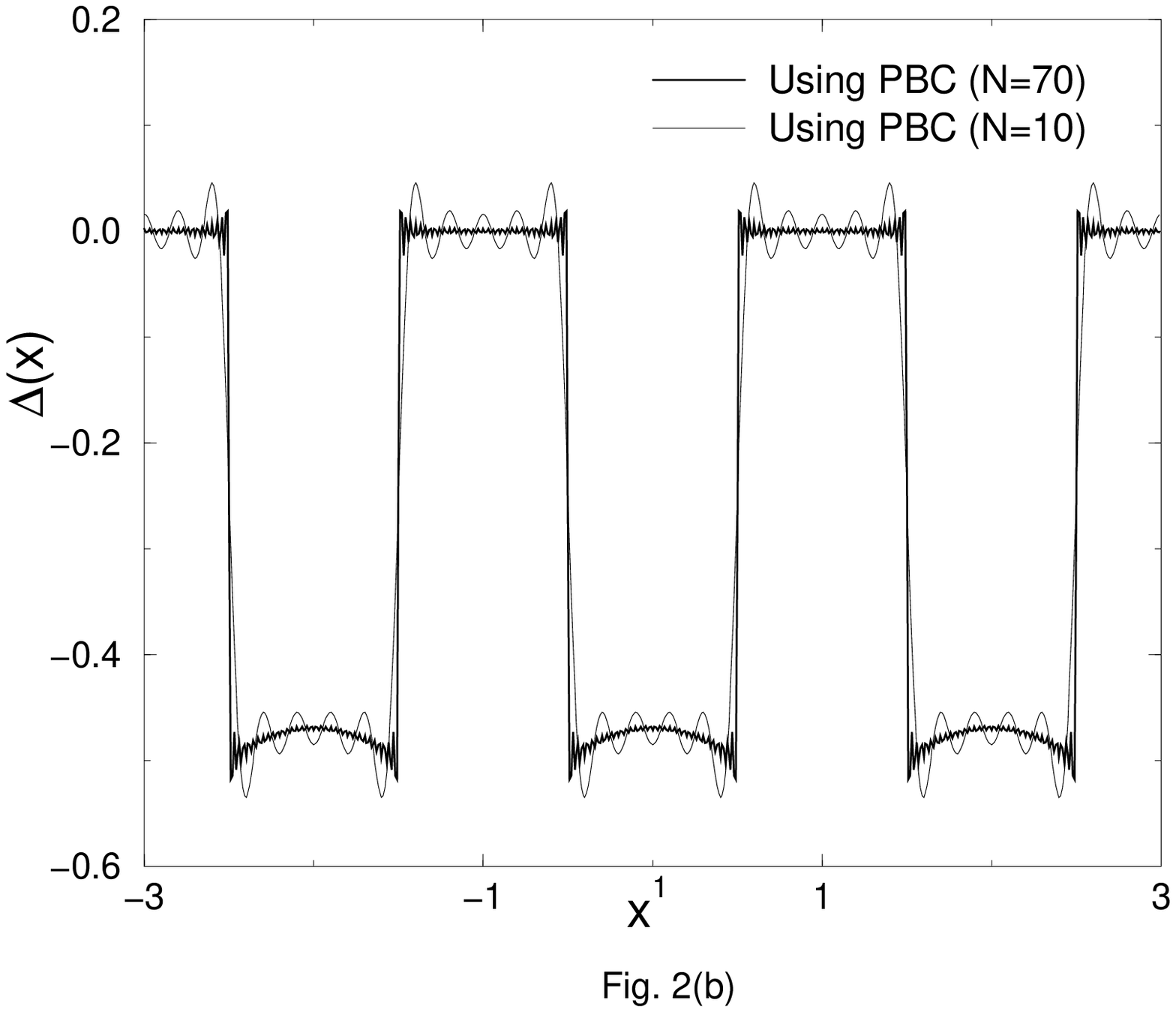,width=5cm,height=7cm}}\ \
\parbox{5cm}{\epsfig{figure=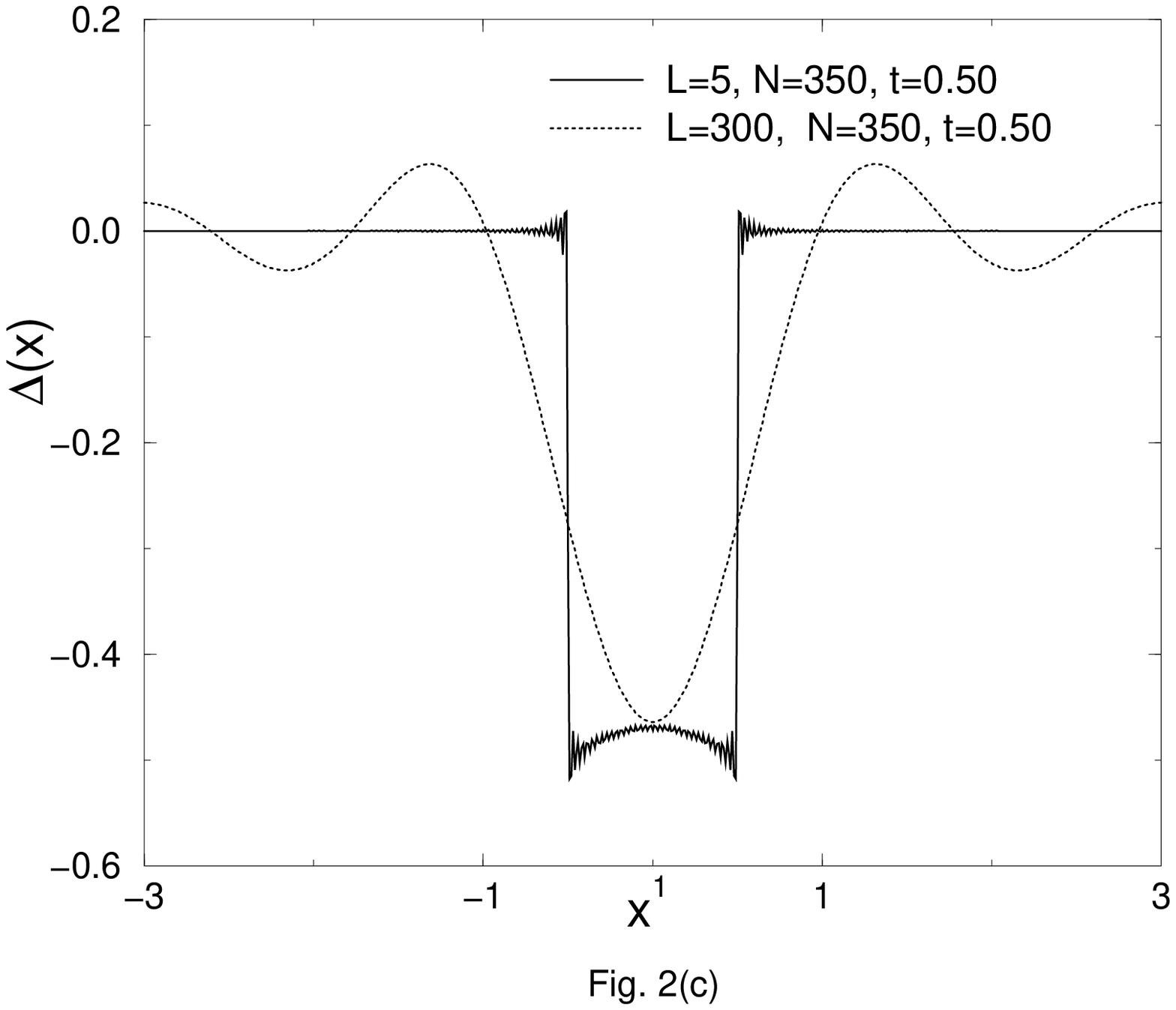,width=5cm,height=7cm}}\ \
\end{center}
\vspace{0.2cm}
\begin{center}
\parbox{14cm}{{\footnotesize 
Fig. 2: Plot of Pauli-Jordan function in equal time. 2(a): continuum
theory for
$x^0=0.5$.
 2(b): DETQ 
theory for $L=1$ and two different values of $N$.
 2(c): DETQ 
theory for $N=350$ and two different values of $L$.}}
\end{center}
\vspace{0.3cm}

This, of course, does not mean that microcausality
is violated. It only means that we are far from continuum limit.
On the other hand,    if
one increases $L$ and $N$ simultaneously keeping the ratio $N/L$
large one
gets the same continuum result within the box. Thus, 
the effect of taking $L$ to
infinity this way simply removes the periodic copies and gives the
continuum result everywhere. 
%It should be noted that the
%ratio $N/L$ for which continuum result is realized depends also on the value 
%of $x^0$ that we are interested in. 

These are, of course, very well known results. The main point that we want to
drive home here is that if we are somewhat  near the continuum limit, then
$\Delta_{ET}$ actually oscillates around the exact result and {\it thus, 
taking an average, one could  obtain good agreement with the continuum result.} In
Fig. 3(a),
we have shown the $\Delta_{ET}$ for a particular space-time point as a
function of $L$. We see clearly that it fluctuates in a very small range
where any value is actually close to the continuum result and hence, the
average value in that range is a good approximation for the continuum limit.
Further, we notice that as $L$ increases $\Delta_{ET}$ converges slowly
but appreciably towards a particular value which is its true continuum
limit. 

In the case of $\Delta_{LC}$, the situation is different. In particular, for
a fixed $L$ the result stabilizes as we increase $N$, but no way near 
result of the continuum limit within the box, as is also observed in
Ref.\cite{sch}. 
\begin{center}
\vspace{0.5cm}
\parbox{7cm}{\epsfig{figure=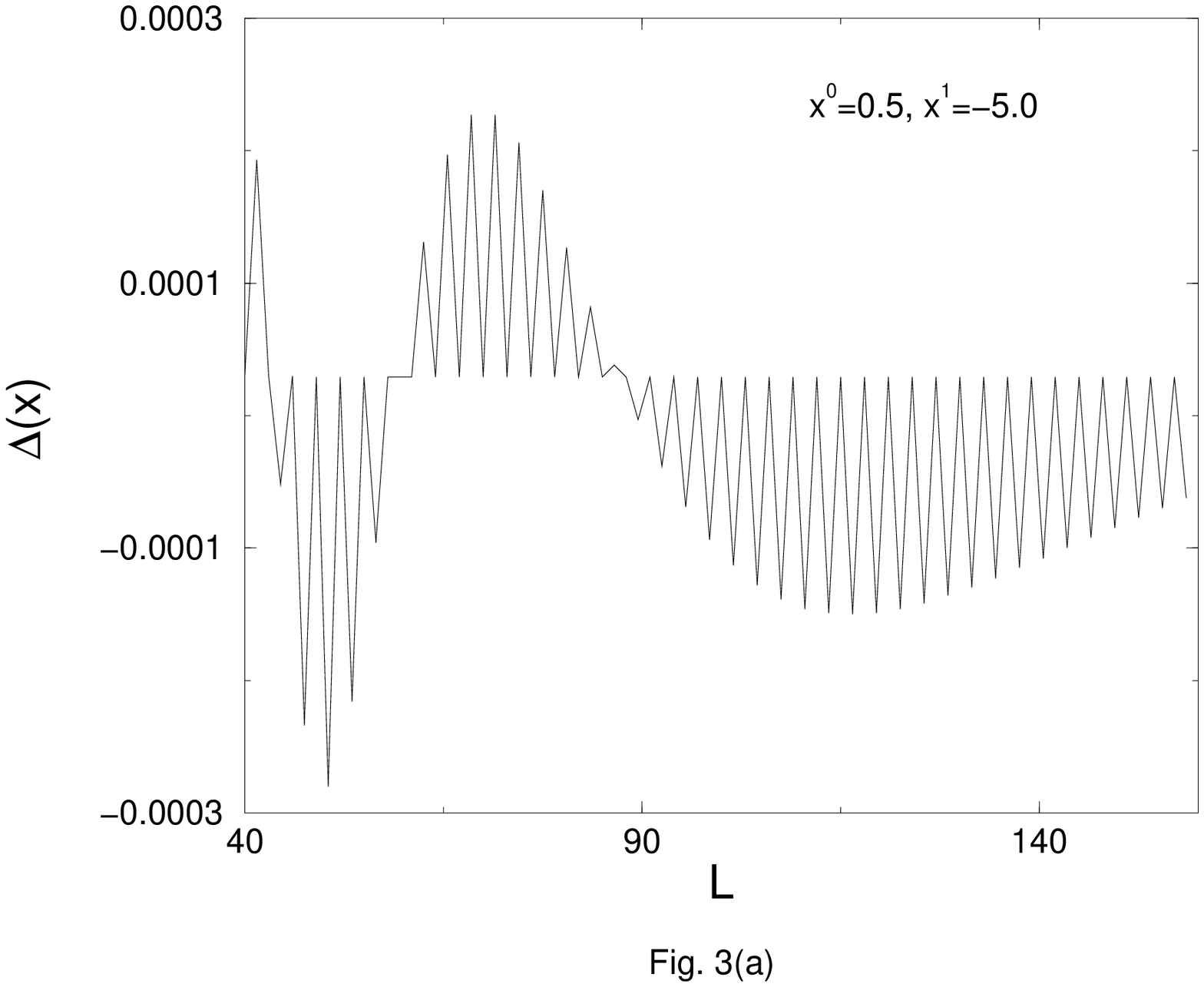,width=7cm,height=6cm}}\ \
\parbox{7cm}{\epsfig{figure=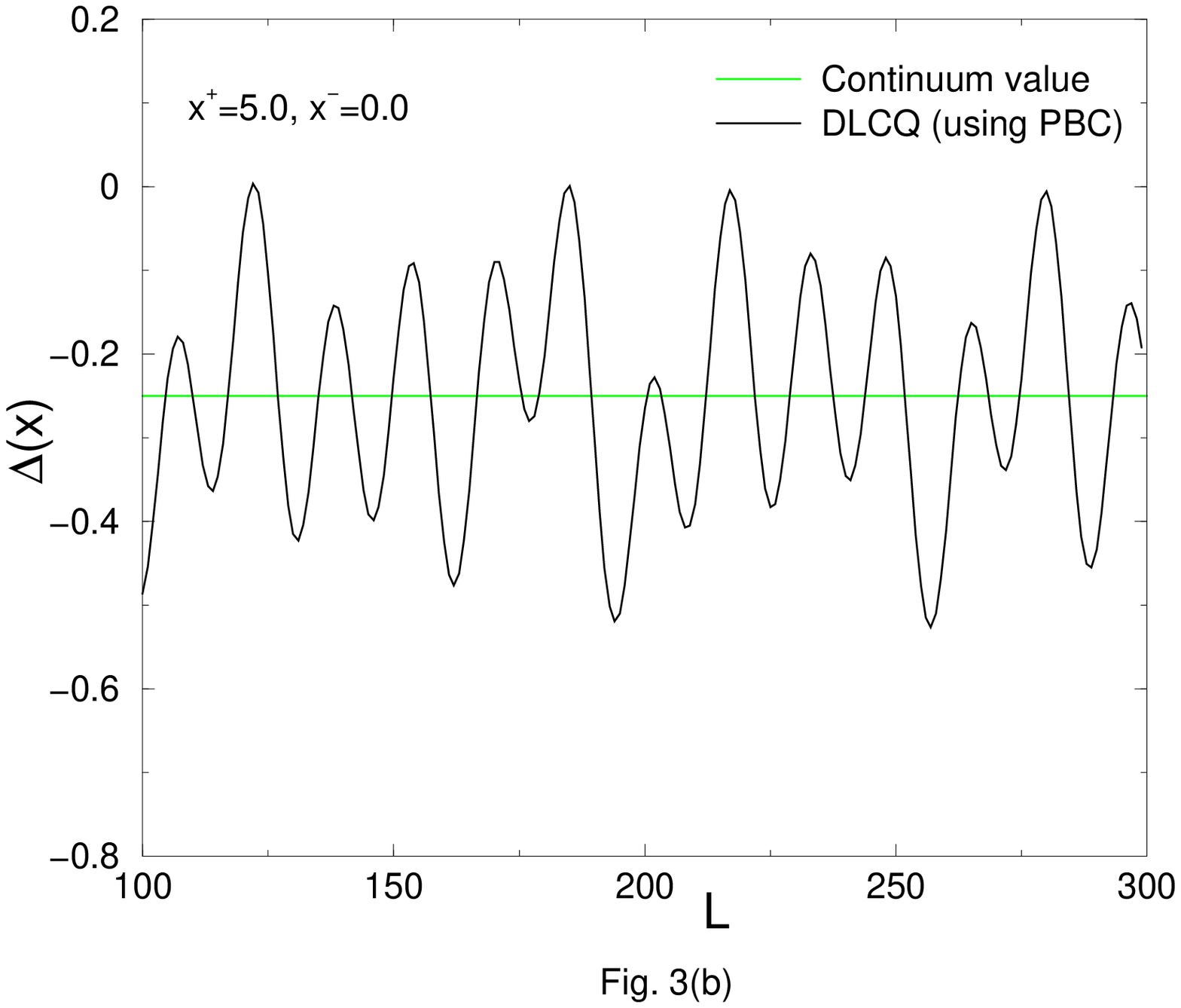,width=7cm,height=6cm}}\ \
\end{center}
\vspace{0.2cm}
\begin{center}
\parbox{14cm}{{\footnotesize 
Fig. 3(a): Variation of $\Delta(x)$ with $L$ in DETQ, 
 3(b): Variation of $\Delta(x)$ with $L$ in DLCQ using periodic
boundary condition.}}
\end{center}
\vspace{0.3cm}

Thus, this result on its own is unphysical 
in the sense that we are
unable to test them by some experiment and it is not surprising to observe
that microcausality is violated (since it is non-zero in the space-like
region). Now, as in Fig. 3(a), we show the similar plot for $\Delta_{LC}$ in
Fig. 3(b).
We clearly see the striking difference in this case, namely, it varies in a
wider range of values and shows hardly any evidence of converging to some
value. Nevertheless, it shares one property of $\Delta_{ET}$, namely, it 
fluctuates around some average value. Taking the clue from earlier plots it 
suggests that the average value about which it is fluctuating might be a good 
approximation for the continuum limit for $\Delta_{LC}$. In fact, it turns
out to be the case. We have shown this fact in Figs. 4(a), 4(b) and 4(c). Here,  
at different $x^+$ and $x^-$, we have first calculated these 
averages for various box length $L(>x^+, x^-)$ with large $N/L$ ratio 
 and plotted them and
the results remarkably agree with the continuum results. Also
 the discretized result is zero in the spacelike region.
  The fact that continuum
result does not care whether we have imposed periodic or anti-periodic
boundary conditions, is also shown in these plots. 

Thus, we see that even
though an individual result with fixed $L$ looks unphysical, an
assembly of such results for various box length $L$ can be used to extract
sensible result in DLCQ and the microcausality, which is required for any
sensible physical result, is in no danger here.  

\vspace{0.3cm}
\begin{center}
\parbox{5cm}{\epsfig{figure=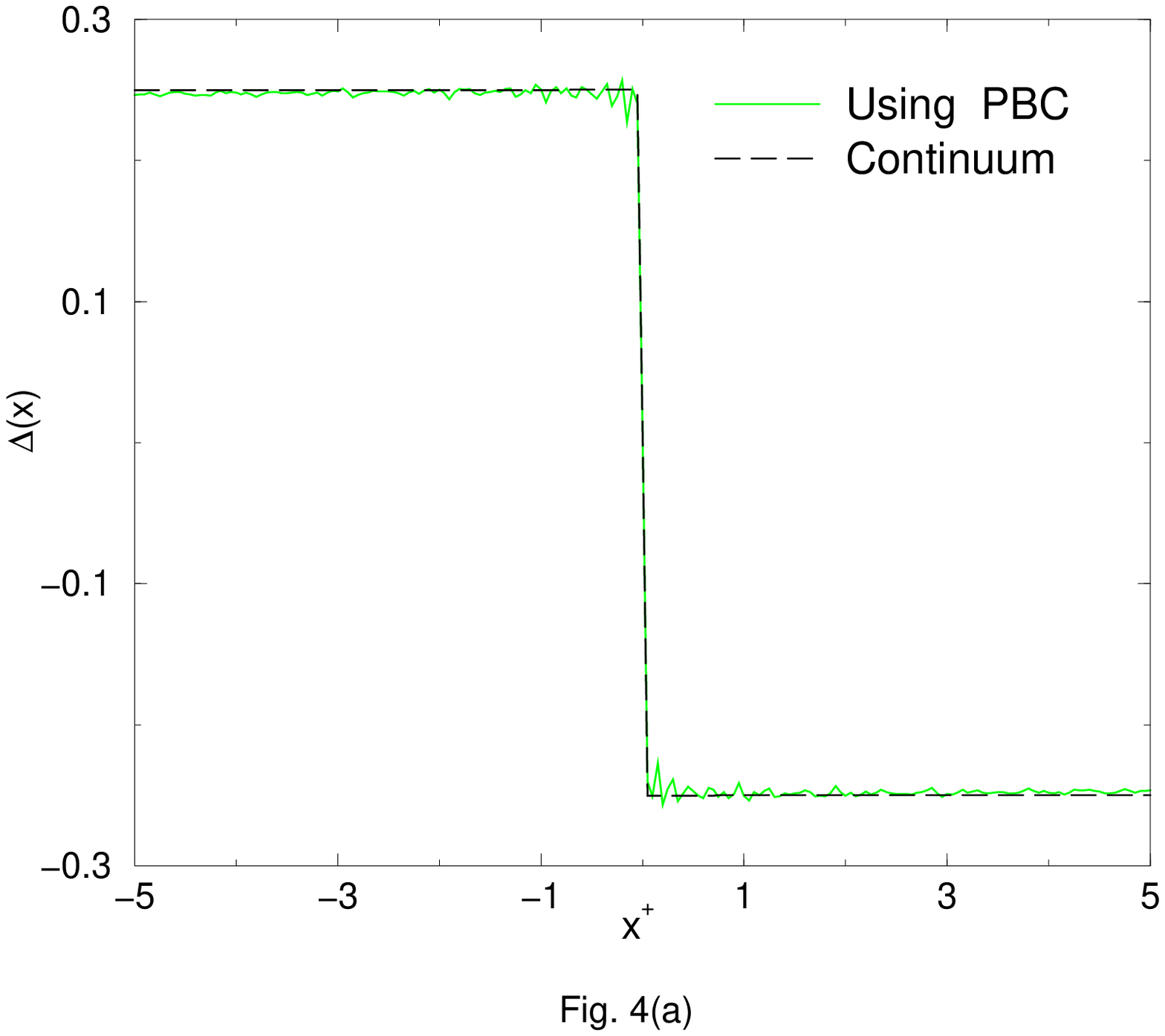,width=5cm,height=6cm}}\ \
\parbox{5cm}{\epsfig{figure=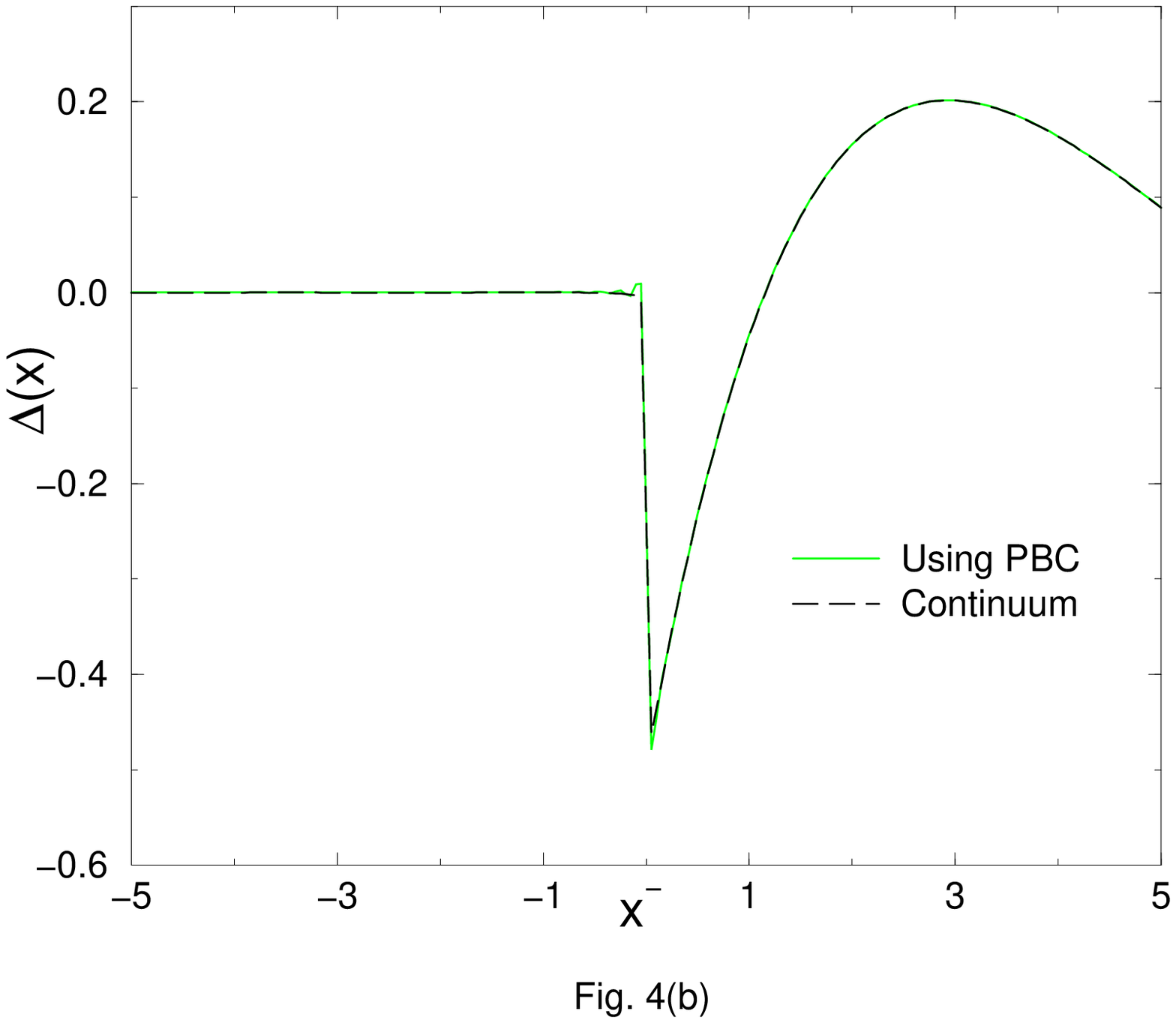,width=5cm,height=6cm}}\ \
\parbox{5cm}{\epsfig{figure=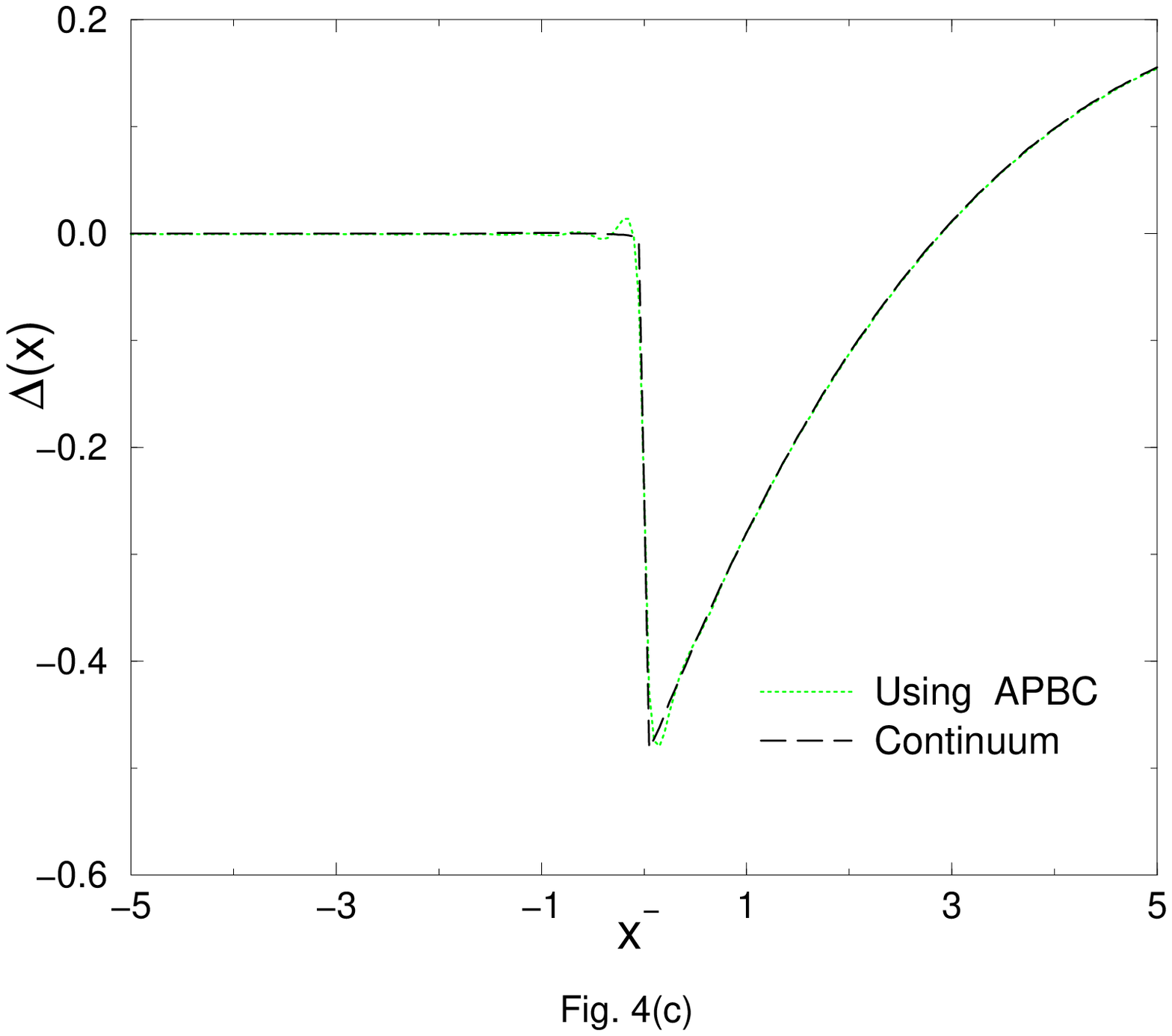,width=5cm,height=6cm}}\ \
\end{center}
\vspace{0.2cm}
\begin{center}
\parbox{14cm}{{\footnotesize 
Fig. 4: Plot of $L$ averaged DLCQ result of $\Delta(x)$ using
 4(a): periodic boundary condition  in comparison
with the continuum result for $x^- = 0$.
 4(b): periodic boundary condition  in comparison
with the continuum result for $x^+ = 5.0$. 
4(c): anti-periodic boundary condition  in comparison
with the continuum result for $x^+ = 2.0$. $\Delta(x)$ is zero in spacelike
region in both 4(b) and 4(c).}}
\end{center}
\vspace{0.2cm}

We have also verified that the same feature is evident in  Feynman propagator
$\Delta_F(x)$ as well. Imaginary part of the Feynman propagator being
connected to the Pauli-Jordan function ($\Delta(x)=2\epsilon(x^+){\rm
Im}(i\Delta_F(x)$), we simply compare the real part of the propagator in this
theory
\begin{equation}
Re(i\Delta_F(x))~=~{1\over 8}[\epsilon(x^+)+\epsilon(x^-)]N_0(m\sqrt{x^2})~-~ 
{1\over 4\pi}[\epsilon(x^+)-\epsilon(x^-)]K_0(m\sqrt{x^2})
\end{equation}
with the corresponding DLCQ result using anti-periodic boundary condition
\begin{equation}
Re(i\Delta_F(x))=\sum_{n=1}^N{1\over 2\pi (2n-1)}~\cos\left( {m^2Lx^+\over
2(2n-1)\pi } + {(2n-1)\pi x^-\over 2L}\right).\label{refy}
\end{equation}  
In Fig. 5, we have shown the DLCQ result which is obtained by using
Eq. (\ref{refy}) for various $L$ and taking averages as we did earlier. 
It is again in
remarkable agreement with the continuum result.

So far we have restricted ourselves to position space. The situation is
completely different if calculations are performed in momentum space where
the continuum limit is easily achieved as discussed in the next section. This
is not surprising as we see that  the Fourier transforms of the DLCQ results
are the discretized versions of the continuum results.  
Fourier transform of $\Delta(x)$ for $x^-=0$ in DLCQ is given by,
\begin{eqnarray}
\Delta(k_j^-)=-{1\over {2\pi}}\int_{-\infty}^{\infty} dx^+ e^{-{i\over
2}k_j^-x^+} \sum_n {1\over {2\pi n}} sin\Big ({k_n^-x^+\over 2}\Big )
= -{1\over {2\pi i}} {1\over k_j^-}.
\end{eqnarray}
It is the  discretized version of the Fourier transform of the continuum result
 (i.e., $-{1\over 4}\epsilon(x^+)$). Similarly, the Fourier transform of $\Delta(x)$ for
$x^+=0$ in DLCQ is also the discretized version of the Fourier transform of
$-{1\over 4}\epsilon(x^-)$.

\begin{center}
\vspace{0.2cm}
\hspace{0.1cm}
\epsfig{figure=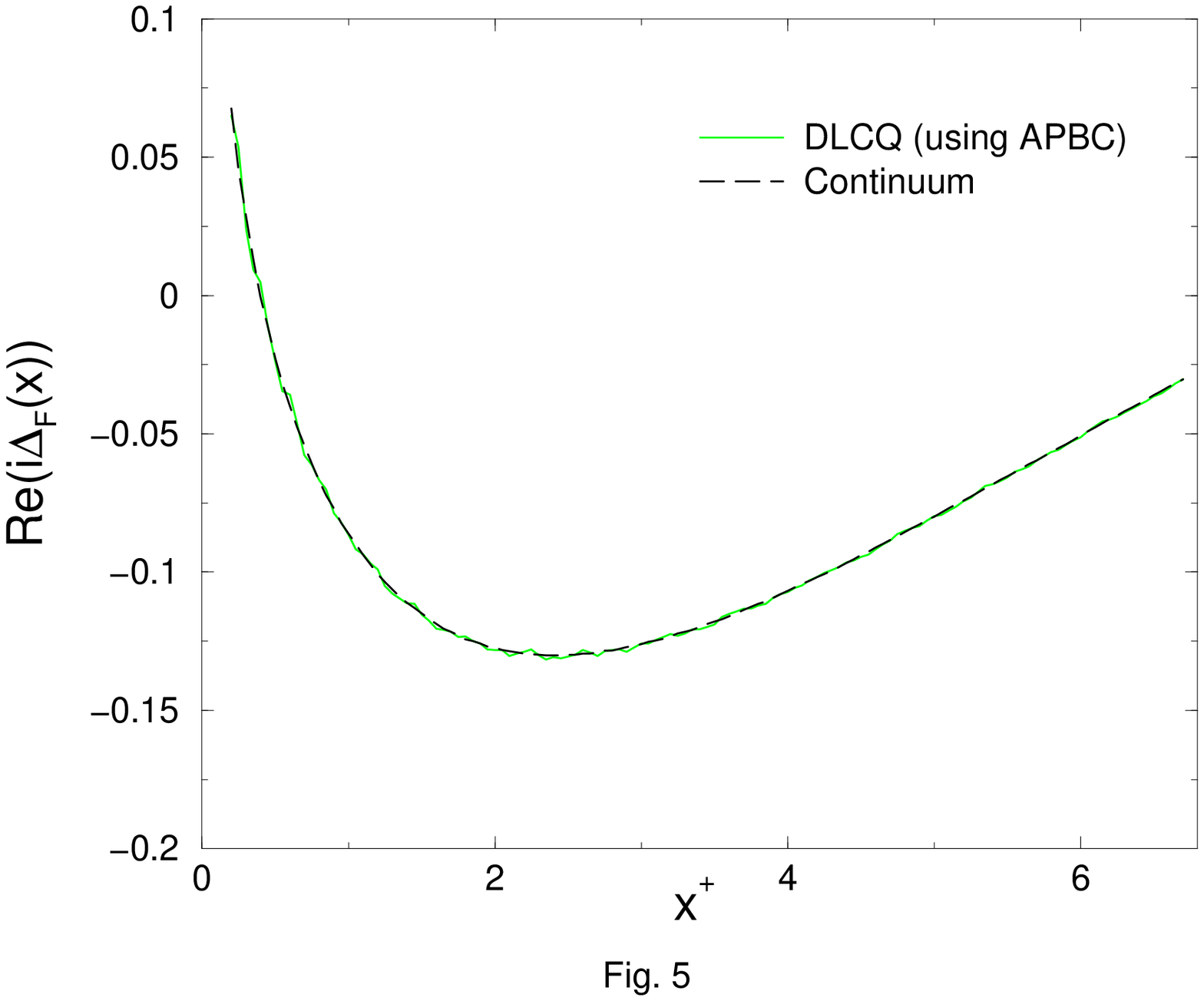,width=6.5cm,height=6.0cm}\\
\end{center}
\vspace{-0.0cm}
\begin{center}
\parbox{14cm}{{\footnotesize 
Fig. 5: Plot of $L$ averaged DLCQ result of 
${\rm Re}(i\Delta_F(x))$ using
anti-periodic boundary condition  in comparison
with the continuum result for $x^- = 2.0$.}} 
\end{center}
\vspace{0.2cm}

%%%%%%%%%%%%%%%%%%%%%%%%%%%%%%%%%%%%%
\noindent{\bf 5. Interacting Theory}
%%%%%%%%%%%%%%%%%%%%%%%%%%%%%%%%%%%%%
\vskip .2in
Till now we have considered Pauli-Jordan function in (1+1) dimensional
free scalar field theory and have shown through numerical analysis that one
gets the continuum result irrespective of using DETQ or DLCQ. In this
section, we investigate the situation by adding a $\phi^4$-interaction. In
particular, we calculate a scattering amplitude in perturbation theory in DLCQ and compare its continuum
limit with the exact result. We shall observe that the 
continuum limit is obtained much more directly 
for such entities which are functions in momentum
space compared to Pauli-Jordan function, which is a function in position
space.
\begin{center}
\begin{picture}(360,120)(0,20)
%\Line(0,60)(400,60)
%\Line(0,60)(0,200)
\SetOffset(0,30)
\SetScale{.7}
\Line(0,60)(30,80)
\Line(0,100)(30,80)
\Line(90,80)(120,100)
\Line(90,80)(120,60)
%\DashLine(60,60)(60,100){1}
\Curve{(30,80)(60,70)(90,80)}
\Curve{(30,80)(60,90)(90,80)}
\Text(110,55)[]{$=$}
\Text(250,55)[]{$+$}
\Text(45,5)[]{$T_{fi}$}
\Text(190,5)[]{$T_{fi(a)}$}
\Text(310,5)[]{$T_{fi(b)}$}
\Text(0,38)[]{$p_2$}
\Text(0,76)[]{$p_1$}
\Text(90,38)[]{$p_4$}
\Text(90,76)[]{$p_3$}
\Text(45,72)[]{$q$}
\SetOffset(140,30)
\Line(0,60)(30,80)
\Line(0,100)(30,80)
\Line(90,80)(120,100)
\Line(90,80)(120,60)
\DashLine(60,60)(60,100){2}
\Curve{(30,80)(60,70)(90,80)}
\Curve{(30,80)(60,90)(90,80)}

%\Text(0,55)[]{$p_2$}
%\Text(0,110)[]{$p_1$}
%\Text(120,110)[]{$p_3$}
%\Text(120,55)[]{$p_4$}
%\Text(55,95)[]{$q$}
%\Text(60,25)[]{6(a)}

\Line(190,70)(300,70)
\Line(190,70)(300,50)
\Line(180,130)(290,110)
\Line(180,110)(290,110)
\Curve{(190,70)(240,100)(290,110)}
\Curve{(190,70)(240,80)(290,110)}
\DashLine(240,130)(240,50){2}

%\Text(170,135)[]{$p_1$}
%\Text(170,100)[]{$p_2$}
%\Text(310,45)[]{$p_4$}
%\Text(310,75)[]{$p_3$}
%\Text(260,80)[]{$q$}
%\Text(240,25)[]{6(b)}
\end{picture}
\end{center}
We consider one loop scattering process in $\phi^4$ theory. The Feynman
diagram shown in the figure is the sum of two time-ordered diagrams.
 The scattering amplitude of the process is given by,
\begin{equation}
T_{fi} = {\lambda^2\over {8\pi}} \int_0^1 dy {1\over { y(1-y)s
-m^2+i\epsilon}}
\end{equation}
where $s=(p_1+p_2)^2$ and $y={q^+\over P^+}$, $P^+$ is the total
longitudinal momentum. We have taken $s < 4m^2$.
In old-fashioned 
light-front Hamiltonian perturbation theory, the second diagram ($T_{fi(b)}$)
in the right hand side is absent.
 DLCQ result for the first diagram can be easily calculated and is given by
\begin{equation}
T_{fi(a)} = T_{fi} = {\lambda^2\over {4\pi}}\sum_{n=1}^N {1\over
{H(2n-1)[K-(2n-1)]-m^2K+i\epsilon}}\label{fdlc}
\end{equation}
We have used the anti-periodic boundary condition here to avoid the zero
mode problem which is present in the interacting theory with 
periodic boundary condition. Here $P^+=p^+_1+p^+_2=K~\pi/L$, $P^-=
p^-_1+p^-_2=H~L/\pi$ , $M^2=P^+P^- = KH$ and
$q_n=(2n-1)~\pi/L$, with $K$ and $n$ being integer. Also, since internal
lines carry momenta which are odd multiple of $\pi/L$, $K$ should be even and
decides the value of $N$, the highest mode in the sum.  
 Note that the result in Eq. (\ref{fdlc}) is
independent of $L$ and the continuum limit is simply obtained by increasing
$N$, (i.e., $K$ here) if it stabilizes. It can be easily verified that 
DLCQ gives the continuum result without any problem as correctly mentioned
in Ref.\cite{yam}. Here we explicitly show that in perturbative $\phi^4$
theory DLCQ in momentum space is
free from zero mode problem and at the same time it gives us a comparative
picture of DLCQ in position and momentum space.  

In old-fashioned equal-time Hamiltonian perturbation theory, contribution
comes from both of the time-ordered diagrams. In this case, DEQT result is
given by  $T_{fi}=   T_{fi(a)} +  T_{fi(b)}$, where,
\begin{eqnarray}
T_{fi(a)} = {\lambda^2\over {16\pi}}&&\sum_{n=-N}^N {1\over {\sqrt
{n^2+{m^2L^2\over \pi^2}}}} {1\over {\sqrt {(K-n)^2+{m^2L^2\over
\pi^2}}}}\Big ({L\over \pi}\Big )^2\cdot \nonumber\\&&~~~~~~~~~~~~~~~~~~
{1\over {[\sqrt {K^2+{sL^2\over \pi^2}}-\sqrt {n^2+{m^2L^2\over \pi^2}}-\sqrt
{(K-n)^2+{m^2L^2\over \pi^2}} + i\epsilon]}}
\end{eqnarray}  
\begin{eqnarray}
T_{fi(b)} = {\lambda^2\over {16\pi}}&&\sum_{n=-N}^N {1\over {\sqrt
{n^2+{m^2L^2\over \pi^2}}}} {1\over {\sqrt {(K+n)^2+{m^2L^2\over
\pi^2}}}}\Big ({L\over \pi}\Big )^2\cdot \nonumber\\&&~~~~~~~~~~~~~~~~~~
{1\over {[-\sqrt {K^2+{sL^2\over \pi^2}}-\sqrt {n^2+{m^2L^2\over \pi^2}}-\sqrt
{(K+n)^2+{m^2L^2\over \pi^2}} + i\epsilon]}}
\end{eqnarray}
The scattering amplitude in this case depends on $L$. However, when the box
length $L$ and the ratio
${N\over L}$ are large the result is independent of $L$ and agrees with the
continuum result. The table below compares the results in some arbitrary
units as obtained by different formalisms. For DETQ, the calculations are
done with $N/L=50$ and $L=5$. But for DLCQ, stability in $K$ 
depends on the mass, for small mass $K$ should be larger than that 
for large mass (e.g., for $m^2=7$, $K\ge 15$, for $m^2=0.1$, $K\ge 80$). For
higher order of accuracy one should take larger values of $N/L$ and $K$.   
\begin{center}
\begin{tabular}{|c|c|c|c|c|} \hline
~~~s~~~~ & ~~~ $m^2$~~~  & ~~~ ${T_{fi}\over \lambda^2}$ (DLCQ)~~~  &~~~
${T_{fi}\over {\lambda^2}}$ (DETQ)~~~  & 
~~~${T_{fi}\over \lambda^2}$ (Continuum)\\ \hline
~15.0  & ~7.0  &  -0.0094    &  -0.0094    &  -0.0094 \\ \hline
~ 0.2  & ~0.1  &  -0.6250    &  -0.6250   &  -0.6250 \\ \hline
~ 3.2  & ~1.0  &  -0.1101    &  -0.1101   &  -0.1101 \\ \hline
\end{tabular}
\end{center}

\vskip .2in
%%%%%%%%%%%%%%%%%%%%%%%%%%%%%%%%%%%%%%%%%%%%%%%%%%%%%%%%%%%%%%
\noindent{\bf 6. Discussion and Conclusions}
%%%%%%%%%%%%%%%%%%%%%%%%%%%%%%%%%%%%%%%%%%%%%%%%%%%%%%%%%%%%%%
\vskip .2in
In this work, first, we considered Pauli-Jordan function 
in (1+1) dimensional free scalar field theory both in equal time and
light-front formulations and their corresponding discretized versions. We
showed that the microcausality (like the boost invariance)  
which is required for any sensible theory
is not guaranteed to be maintained in the discretized
version of the theory. In practice, 
one is actually interested in the continuum limit of the results 
obtained in discretized version of the theory, which one can presumably test
in some experiment. In general, one expects to obtain the continuum result
by taking $N\rightarrow\infty$ and $L\rightarrow\infty$. 
In DETQ, in coordinate space, for sufficiently large value of $N$ and for any
fixed $L$, continuum result is reproduced within the box as long as 
$x^0\le L$.  Thus taking $L\rightarrow\infty$ limit only removes the
periodic copies. In contrast, same 
procedure does not give the continuum result 
even within the box in DLCQ. Nevertheless, we observe that as a function of
$L$, DLCQ result actually fluctuates around the continuum result even when
$L$ is very large. We noted the fact that taking an average value from an
assembly of such values obtained with various different $L$ reproduces the
continuum result with remarkable agreement. We have also obtained the
continuum limit of the real part of the propagator starting from DLCQ result.
Thus obtaining continuum result ensures microcausality as well as boost
invariance. 

Note that the Pauli-Jordan function or the Feynman propagator are
considered to be the functions in
position space here. In contrast, we showed with a specific example 
that in the perturbative
calculations of diagrams in interacting theory where calculations are
performed in momentum space, DLCQ result becomes independent of $L$ and the
continuum result is obtained simply by increasing $N$ sufficiently.
%%%%%%%%%%%%%%%%%%%%%%%%%%%%%%%%%%%%%%%%%%%%%%%%%%%%%%%%%%%%%%%%%%%%%%%%%%%%%

%%%%%%%%%%%%%%%%%%%%%%%%%%%%%%%%%%%%%%%%%%%%%%%%%%%%%%%%%%%%%%%%%%%%%%%%%%%%%

\end{document}